\documentclass[a4paper,12pt]{article}

\usepackage{graphicx}
\usepackage{amsmath}
\usepackage{amsfonts}
\usepackage{amssymb}
\usepackage{fancyhdr}
\usepackage{epsfig}
\usepackage{float}
\usepackage{array}
\usepackage{times}
\usepackage{wrapfig}

\def\AVE#1{E\left[ \, #1 \, \right]}
\def\AVEX{\overline{A}}
\def\STDX{\Delta{A}}
\def\VARX{\STDX^2}
\def\VAR#1{E\left[ \left(\, #1 \, \right)^2 \right]}

\begin{document}

\title{El Farol Revisited}
\author{Hilmi Lu\c{s}\thanks{\textit{Corresponding author. e-mail:}
hilmilus@boun.edu.tr}, Cevat Onur Ayd{\i}n, Sinan Keten, \\
Hakan \.{I}smail \"{U}nsal, Ali Rana At{\i}lgan \\
\textit{Bo\u{g}azi\c{c}i University, School of Engineering,
\.{I}stanbul, Turkey}}

\maketitle

\begin{abstract}

This article is concerned with the global behavior of agents in
the El Farol bar problem. In particular, we discuss the global
attendance in terms of its mean and variance, and show that there
is a strong dependence of both on the externally imposed comfort
level. We present a possible interpretation for the observed
behavior, and propose that the mean attendance converges to the
\textit{perceived} threshold value as opposed to the actually
imposed one.

\end{abstract}

\newpage

\section{Background and Problem statement}

Modeling inductive reasoning and bounded rationality was the
subject of the influential article by Arthur\cite{arthur94},
wherein the author investigated the global behavior of a group of
agents that independently decided, at each week, whether to go to
a bar. This prototype of agent based modeling, which became known
as the El Farol bar problem, has since been the subject of many
studies, either in its original form or in the context of the
Minority Game (see, e.g., the works by Challet and
Zhang\cite{cz97,cz98}, Johnson et al.\cite{jjjckh01}, Challet et
al.\cite{cmo03}, Cara et al.\cite{cpg99}, and the references
therein).

Here we go back to the original form of the El Farol bar problem
and discuss some issues regarding the global behavior of the
agents in terms of how they perceive the imposed comfort levels.
Our computer experiments simulate the following situation: At each
week $t$, $N$ people independently decide whether to go to the
bar. Their decisions are based on their individual forecasts of
attendance for that week, such that if their estimate is greater
than a specified threshold value $T$ they refrain from going
(discomfort), whereas they visit the bar if they expect less
people than $T$ to show up. Each agent (person) chooses her
estimator from a set of $N_P$ number of predictors. If the
predictor they used on a specific week leads them to take the wise
decision (i.e., she goes and finds the attendance to be indeed
less than $T$, or she does not go and the attendance turns out to
be greater than $T$), then they use that same predictor for the
following week as well; otherwise, they randomly choose another
predictor from the available set. The whole past time history of
attendance is available to all the agents.

The setup we investigate is very similar to those previously
discussed in the literature (e.g. the experiments discussed by
Arthur\cite{arthur94}, Johnson et al.\cite{jjjckh01}, and Challet
et al.\cite{cmo03}, among others), with either all or some of the
following differences: \textit{(i)} The agents are free to choose
their predictors from the whole pool as opposed to each having a
smaller set of her own. Our preliminary investigations have
suggested that these two approaches lead to qualitatively similar
results. \textit{(ii)} Our predictors employ various schemes that
include arithmetic and weighted averages of various
consecutive/non-consecutive weekly data, mirror images, curve
fittings using various number of previous data, observed min/max
values, same value as that of a previous week, etc. Each predictor
may require a different length of past data. \textit{(iii)} Each
agent uses her most recent predictor as long as it leads to wise
decisions, and randomly chooses a new one when it fails; in other
words, the agents do not keep records of past performances of the
predictors expect for the most recent one. \textit{(iv)} The
initial attendance data required to start each simulation is
provided randomly. We have observed that the steady state behavior
of the group is independent of these initial conditions, provided
that the predictors are able to represent a wide variety of
behavior patterns.

The main question we aim to discuss is: Is the collective behavior
of the agents affected by the comfort level, i.e. the ratio $T/N$?

\section{Behavior for various thresholds}

We use a pool consisting of $76$ predictors, and analyze the
problem for various values of $N$ and $N_P$ people. We consider
both $1000$ and $5000$ time steps in the simulations; we should
note, however, that the statistics do not differ significantly as
we extend the time horizon. We denote by $A(t)$ the time history
of attendance (and also the value of attendance at week $t$,
depending on the context), and try to quantify the collective
behavior in terms of the sample average $\AVEX=\AVE{A(t)}$ and the
sample variance $\VARX=\VAR{A(t)-\AVEX}$ where $\AVE{ \, \cdot
\,}$ denotes the expected value operator.

\begin{figure}[p]
\center

\epsfig{figure=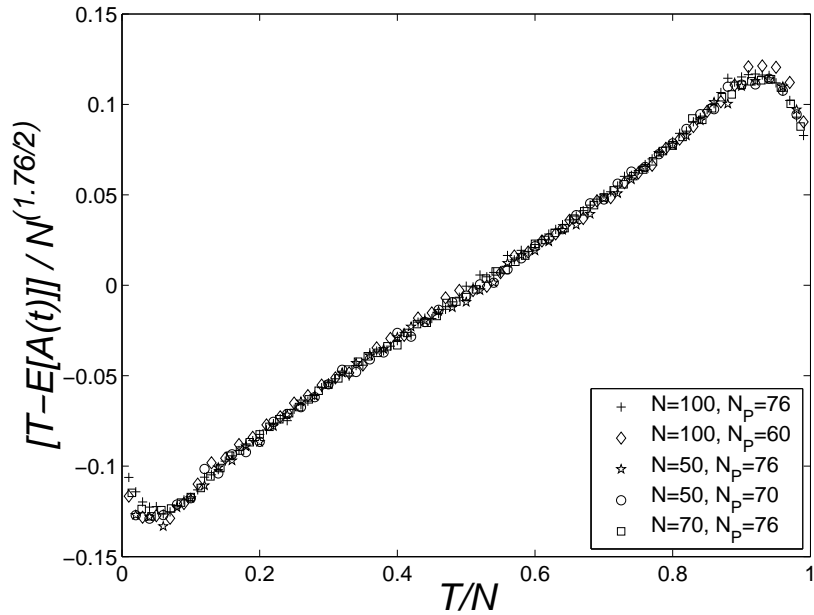,width=0.7\textwidth}
\caption{Deviations of observed means from the threshold values
for various values of the ratio $T/N$.}\label{TminusAvervsTN}

\end{figure}

Figure \ref{TminusAvervsTN} shows that the global attendance has a
mean near, but not equal to, $T$. The deviations are a function of
the ratio $T/N$; moreover, the relationship is almost perfectly
linear except for the end regions $T/N < 0.2$ and $T/N > 0.8$. It
is noteworthy that the mean converges to the threshold only when
$T/N=1/2$; whenever $T<N/2$ the mean exceeds $T$, and whenever $T
> N/2$ the mean is smaller than $T$. Note that the scaling is
$1/N^{(1.76/2)}$ for the vertical axis: This scaling turns out to
be (intuitively) necessary since the absolute deviations depend on
the number of people. What is perhaps more interesting is that the
scaling that yields the best overlap is not $1/N$ but rather some
other power of it.

\begin{figure}[p]
\center

\epsfig{figure=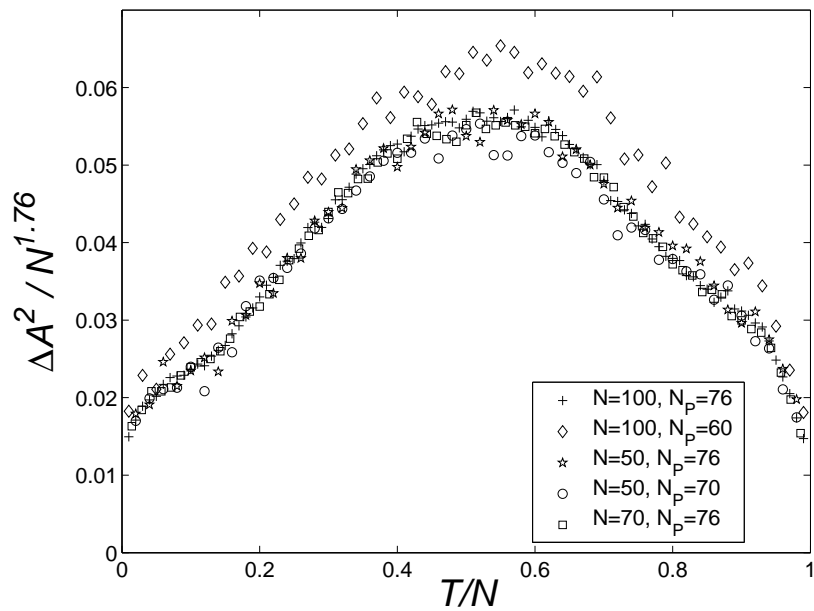,width=0.7\textwidth}
\caption{Variances of global attendance time histories
  for various values of the ratio $T/N$.}\label{VARvsTN}

\end{figure}

In Figure \ref{VARvsTN} we present the sample variances as a
function of $T/N$. For all cases considered, the curves are
suggestive of a quadratic relation. Here the effects of the number
of predictors used may be observed in the outlier results for
$N=100$ people and $N_P=60$ algorithms. Our simulations led us to
conclude that as long as the number of predictors used is not
(significantly) small when compared with the number of people,
$N_P$ has no pronounced effect neither on $[T-\AVE{A(t)}]$ vs.
$T/N$ nor on $\VARX$ vs. $T/N$. When, however, $N_P$ is
significantly small, then there form somewhat large groups of
people that use the same predictor and behave in the same way.
These groups, in turn, increase the variance, such that the
smaller the number of predictors, the larger the number of people
that behave identically and the larger the variance. The limit, of
course, is $N_P=1$ for which we would have $\VARX=N^2/4$.

Finally, we strongly emphasize that we can not claim any physical
significance for the number $1.76$; as it is, it is just a scaling
factor we use, which leads to patterns that are visually
overlapping. Our numerous simulations suggest that keeping $N$
constant and varying $N_P$ or vice versa (subject to
aforementioned conditions) does not alter the value of this
constant, as implied by Figures \ref{TminusAvervsTN} and
\ref{VARvsTN}.

\section{Discussions and Conclusions}
These global patterns are somewhat surprising in light of some
previous comments in the literature \cite{arthur94,cmo03,cpg99},
except for the analysis by Johnson et al. \cite{jjjckh01} wherein
the authors mention that the mean attendance is near, but not
equal to, the threshold. We now try to formulate an interpretation
for the observed phenomena as follows: Let $U(A(t))$ denote some
(unknown) function which leads to
\begin{equation}
\AVE{U(A(t))}= U(T) \label{utility1}
\end{equation}
We immediately note that for $\AVEX=T$, the function $U$ should be
linear. This is a macro (global) condition, and as it is it can
not be imposed directly on the agents (micro), even though the
threshold itself is imposed globally. Therefore the convergence of
$\AVEX$ to $T$ is not guaranteed. Our results in fact suggest that
the system as a whole adjusts itself so as to impose a
\textit{perceived} threshold to which $\AVEX$ converges. To dwell
further into the details of this behavior, let us expand both
sides of eq.\eqref{utility1} about $\AVEX$ and keep only the small
order terms to get
\begin{equation}\label{utility2}
\AVE{U(\AVEX) + [A(t)-\AVEX]U'(\AVEX) +
\frac{[A(t)-\AVEX]^2}{2}U''(\AVEX) } = U(\AVEX) +
[T-\AVEX]U'(\AVEX)
\end{equation}
which leads to
\begin{equation}
[T-\AVEX]= \frac{\VARX}{2} \frac{U''(\AVEX)}{U'(\AVEX)}
\label{utility3}
\end{equation}

\begin{figure}[p]
\center

\epsfig{figure=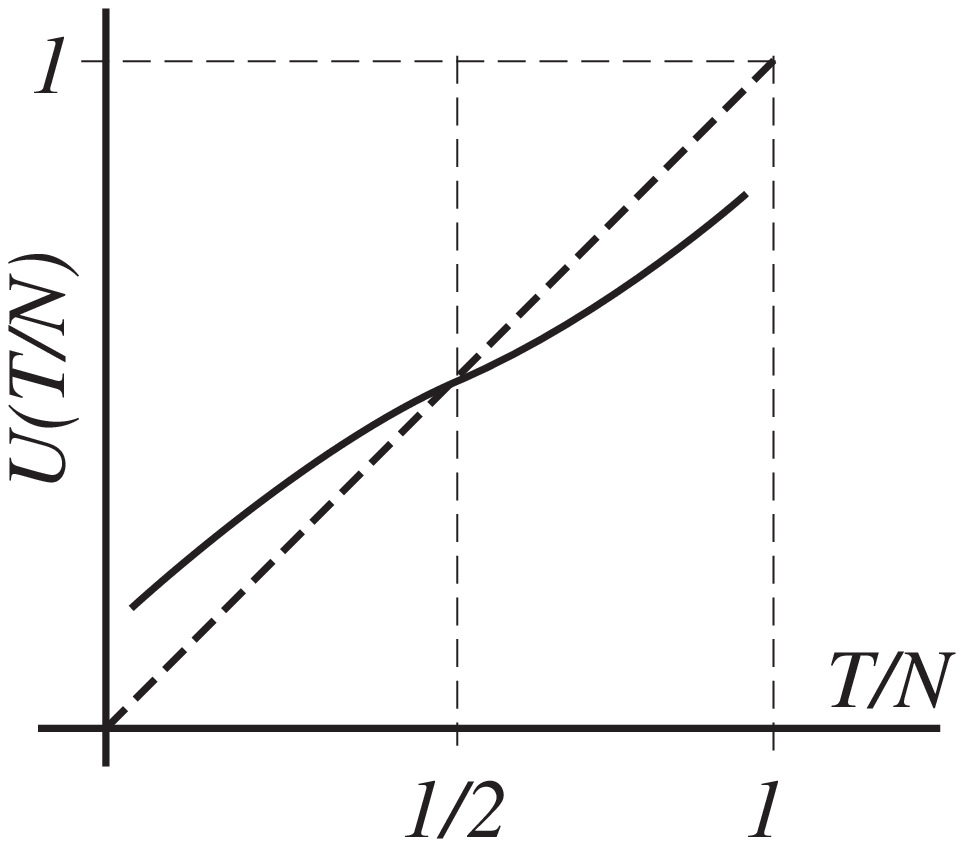,width=0.4\textwidth}
\caption{Global perception of the externally imposed threshold as
a function of $T/N$.}\label{PerceptionFunction}

\end{figure}

The expansion given in eq.\eqref{utility2} is reminiscent of the
approach discussed by Pratt\cite{pratt64}. Provided that $U$ is a
monotonically non-decreasing function, the sign of the right hand
side of eq.\eqref{utility3} is governed by the sign of
$U''(\AVEX)$; $[T-\AVEX] < 0$ implies a concave $U(\AVEX)$, and
$[T-\AVEX] > 0$ implies a convex $U(\AVEX)$. We could interpret
the observed behavior by focusing on the relation of the global
attendance with respect to the applied threshold, and say that
mean attendance converges to the perceived threshold instead of
the actually imposed value, and that the global perception is
defined by a function $U$ of the form shown in Figure
\ref{PerceptionFunction}.

\begin{thebibliography}{99}

\bibitem{arthur94} W.B. Arthur. \textit{Inductive reasoning and bounded rationality}.
Amer. Econ. Rev. \textbf{84}, 406, 1994.

\bibitem{cz97}D. Challet and Y.-C. Zhang. \textit{Emergence of cooperation and organization in an evolutionary
game}. Physica A \textbf{246}, 407, 1997.

\bibitem{cz98}D. Challet and Y.-C. Zhang. \textit{On the minority game: Analytical and numerical studies}.
Physica A \textbf{256}, 514, 1998.

\bibitem{jjjckh01} N.F. Johnson, S. Jarvis, R. Jonson, P. Cheung, Y.R.
Kwong, and P.M. Hui. \textit{Volatility and agent adaptability in
a self-organizing market}. Physica A \textbf{258}, 230, 2001.

\bibitem{cmo03} D. Challet, M. Marsili, and G. Ottino. \textit{Shedding light on El Farol}.
Physica A, \textbf{332}, 469, 2003.

\bibitem{cpg99} M.A.R. de Cara, O. Pla, and F. Guinea.
\textit{Competition, efficiency, and collective behavior in the
``El Farol'' bar model}. Eur. Phys. J. B \textbf{10}, 187, 1999.

\bibitem{pratt64} J.W. Pratt. \textit{Risk aversion in the small
and in the large}. Econometrica \textbf{32}, 122, 1964.

\end{thebibliography}
\end{document}